# Is Dissipative Granular Gas in Knudsen Regime Excited by Vibration Biphasic ?

**P. Evesque**
Lab MSSMat, UMR 8579 CNRS, Ecole Centrale Paris
92295 CHATENAY-MALABRY, France, e-mail: evesque@mssmat.ecp.fr

**Abstract:**

*Investigation is pursued of the simple model I have proposed recently to describe a granular gas in Knudsen regime, in micro-gravity, and excited by a vibrating piston (vibration direction along Oz according to z=bω cos(ωt)). This model predicts a probability distribution function f(v) of speed v in the direction of vibration whose tail varies approximately as $f(v) \propto (1/v) \exp(-v/v_o)$, for which $v_o$ obeys $v_o=\beta b\omega/(\alpha n_l)$; here $n_l$ is the number of granular layers in the cell at rest and β/α is a constant coefficient whose range is $0.06< \beta/\alpha <2/3$. This model results from a specific non local coupling between dissipation that occurs during a roundtrip due to ball-ball collisions and speed amplification due to ball-piston collision. It explains the main trends of the distribution p(I) of impacts I with a fix target. This trend has been obtained in recent experiments in board of the Airbus A300-0g of CNES, giving $p(I) \propto \exp(-I/I_o)$. It predicts also the rate $N_c$ of collisions with a fix gauge perpendicular to vibration; it finds $N_c$ varies linearly with the gauge surface S and with the piston speed bω, but is independent of the number N of balls; the theory leads to a correct estimate of the experimental $N_c$. However, as the experimental $N_c$ depends slightly on N, a second phase of balls "nearly at rest" is assumed to exist in order to explain the $N_c$ vs N dependence. This phase describes balls "merely at rest", which are in excess compared to the f(v) prediction; the dependence of $v_o$ on this second phase is discussed. Compatibility between results from granular gas experiments in micro-gravity and experiments on Maxwell's demon in 1-g is discussed. The main result of the paper is that the probability distribution function of speed v along z, i.e. $f(v) \propto (1/v) \exp(-v/v_o)$, diverges as 1/v at small speed and is quite non Boltzmannian at large speed. Hence this makes the granular gas in Knudsen regime a peculiar problem, completely different from classic statistical mechanics. The main idea which allows understanding these results is to consider the piston playing the role of an impact generator or of a "velostat" instead of a thermostat. It is shown also that the model predicts completely different behaviour in 1g.*

**Pacs # : 05.45.-a, 45.50.-j, 45.70.-n, 81.70.Bt, 81.70.Ha, 83.10.Pp**

Many papers ([1-3] and refs there in) have been dealing with dissipative granular gases recently, because these systems are expected to exhibit rather unusual behaviours. On the other hand, many of them start settling the problem using concepts from classic non dissipative systems, such as temperature, …., which may not be adapted. Most of them also neglect analysing or determining the role plaid by boundary conditions. However, we know from recent experiments how non extensive the physics of granular dissipative gas is. In recent experiments granted by CNES and ESA in microgravity on a "gas" of balls confined in a closed fix cylindrical cell and excited by a vibrating piston, which moves according to b cos(ωt) in the direction of the cell axis, our team have found a series of anomalous behaviours [4, 5]. I have attributed these results to specific boundary conditions and to local dissipation [5] and proved that the modelling predicts a strongly non Boltzmannian distribution [6] which agrees with the experimental one.

Indeed, the study of a granular gas in a Knudsen regime [5-6] shows how far the speed distribution of such a dissipative gas is from Boltzmannian. It comes from the dynamics which results from the coupling between (i) a specific boundary effect, which forces the system, and (ii) specific coupling between propagation and dissipation processes. It results in a predicted speed distribution $f(v) = (A/v) \exp(-v/v_o)$, where v is the speed in the direction of vibration. This behaviour is found experimentally with the right dependence of $v_o$ on the number of balls contained in the cell, *i.e.* $v_o \approx 1/N$, [6]. The effect of a series of different possible experimental bias has been also discussed in [6], such as the non linear response of the gauge,…

In the first part of the present paper the description and the prediction of the model are pursued; then they are compared to the experiments. It is shown in particular that the number $N_c/T_{measure}$ of collisions per unit of time with the gauge is predicted to be independent of the ball number N, and to be linearly dependent on both the excitation speed $b\omega$, and on the surface area S of the cell or of the gauge, in the limit $b/L \ll 1$ and if all the balls are in the "gas". In fact the experiment shows that $N_c$ is a continuously increasing function of N, in the experimental range (number of layer smaller than 1.3). This is why the paper discusses the case when all the balls are not in the same "gas" phase, but separate into two phases, the "gas" phase and a second phase where the balls are merely at rest. Both phases are intermixed and interact with each other forming a dynamical equilibrium.

In fact the definition of the second phase requires at first to investigate and define in more details the gas phase itself. Indeed, the "gas" distribution f(v) is $f(v) = (A/v) \exp(-v/v_o)$; so it exhibits a **diverging probability of getting particles** with very slow speeds. Hence, it requires a cut-off speed $v_1 \ll b\omega$ at slow speed adapted to the "normalisation" factor A; both A and $v_1$ ensure the normalisation of f(v) to be $N_g/N \leq 1$. This fixes a relation between A, $v_1$ and $N_g/N$. Here $N_g$ is the number of particles (among the N) which behave as a "gas" at a same time, *i.e.* that obey the anomalous statistics $f(v) = (A/v) \exp(-v/v_o)$. This distribution is obtained under the condition that any particle which hits the piston gains the same amount of speed $\beta b\omega$.

As a matter of fact, slow balls are hardly excited by direct collision with the piston because this occurs when piston is at maximum elongation; so there might be some excess of slow balls. This generates cases for which $N_g < N$. On the other hand, slow balls can arrange spatially due to (i) mutual interaction, or (ii) to interaction with the walls, or (iii) interaction with balls in the gas; this is expected to occur when clustering forms or if the balls with little speed organise themselves in lines parallel to vibration; another case might be similar to the organisation of natural heterogeneity such as in the universe, which is formed of stars, clusters of stars, galaxies, clusters of galaxies…. So it might occur that collision between balls from the "gas" and balls nearly at rest is rarer than normal. In this case one expects $v_o$ to be larger than predicted. This is discussed by introducing a gas phase and a phase "nearly at rest".

Global equilibrium relations linking the relative concentration of the two phases are described as a function of $N_c$ and N. So this is the second part of the paper, which discusses on the utility/necessity of speaking in terms of a two-phase system. It uses recent results of impact distribution in the direction of vibration.

One may ask also if the gas phase, which is defined as the phase whose speed distribution is $f(v) = (1/v) \exp(-v/v_o)$, is equivalent to a skin effect which disconnect the inner part to the boundary condition. This cannot be true according to the present modelling which shows that the balls in the gas phase perform permanent roundtrip from a boundary to the opposite one. So this gas phase fills up and sounds the complete cell.

The last part of the paper tries and investigates if these data, and the theory which explain them, are compatible with data obtained from other experiments such as the Maxwell's demon experiment. In this experiment for instance, one observes a dynamic equilibrium formed in between two half-boxes connected by a slit. It measures the flow of balls perpendicular to the vibration. Experimental results show that the transfer rate decreases as soon as $n_1<0.5$ about in this case. The collisions measured in the direction parallel to vibration show that $N_c$ is still increasing for $n_1>1$. Are these two results compatible?

## 1. Distribution of fast balls in dissipative gas in Knudsen regime :

The model used in the paper has been first stated in p. 28 of [5], then modified and discussed in [6]. It considers an assembly of N identical balls of diameter d and mass m contained in a cylindrical cell of section $S=\pi D^2$ and length $L_o=L+d$. The cell is closed on top by a wall (fix in the cell frame) and on bottom by a vibrating piston, which moves at frequency $f=\omega/(2\pi)$ in the z direction according to $z_p=z_{po}+ b \cos(\omega t)$. so the cell axis defines Oz; be $\varepsilon_b$, $\varepsilon_p$ and $\varepsilon$ the normal restitution coefficients, *i.e.* ball-ball, ball-piston and ball-wall collisions respectively (They will be stated equal, except when precised). One defines also $n_l=Nd^2/D^2$ and the typical mean free path $l_c = D^2L/(4d^2N) = L/(4n_l)$; $n_l$ corresponds approximately to the number of ball layers covering the piston at rest in 1g. In the following v will correspond to the ball speed in the z direction (of vibration) and $\mathbf{u}=(u_x, u_y)=(u_r,u_\theta)$ characterises the ball speed perpendicular to z. The model is interested in the speed distribution f(v) dv and in the distribution p(I) of impacts intensity I on the top wall. This last one will be assumed to be proportional to the momentum transfer in the z direction during the collision, *i.e.* $I=mv(1+\varepsilon)$.

The model states that the gas is a dissipative medium which slows down efficiently the speed of an incident speedy ball. So the only way to get a speedy ball with a large speed $v \approx k\, b\omega$ (k>1) is to find a ball that has flown across the cloud few (k') times without interacting with it; so it has hit k' times the piston successively without hitting any other ball of the gas. The probability g(v) of such an event is small and varies as $\{\exp[-2(L_o-d)/l_c]\}^{k'}= \exp[-2k'L/l_c]$, where $l_c$ is the mean free path between two ball-ball collisions. As $l_c = L/(4n_l)$, the probability g(v) of getting the speed v is $g(v)=\exp[-2k'L/l_c]= \exp[-8k'n_l]$.

As a matter of fact, the factor 8 in g(v) comes from the factor 4 in $l_c$; however it is a maximum estimate, since very eccentric collisions are likely inefficient. Hence, it is better introducing a factor α<8 instead of 8 in g(v) and writes g(v) = exp[-αk'$n_l$] .

It remains to evaluate k'. Indeed, let us assume that each time the ball hits the piston, it gains some amount of speed proportional to bω, *i.e.* βbω ; hence k'=v/(βbω), which leads to g(v)=exp[-αv$n_l$/(βbw)]. In practice, one expects β ≈ 0.5-1-or-2: if the ball is (i) quite speedy (v>>bω), or (ii) quite slow (v<<bω) the average gain is small, *i.e.* β<<1, since the ball either (i) hits the piston with merely equal probability at any moment when it goes back or forth (this corresponds to v>>bω) so that the mean gain, obtained after averaging over all possible phases, is small, or (ii) the ball hits the piston at its maximal elongation only, when its speed is slow, v<< bω, and the gain is small too. At last, the maximum gain cannot overpass β=2(1-ε) , *cf.* [7].

In fact g(v) is the probability that the situation described earlier occurs, without taking into account the time it lasts. Hence, the configurations have to be time averaged to get the real probability. The transient state at speed v lasts a time $T_v$ equal to the travel time for a piston-to-piston roundtrip; hence $T_v$=2L/v . This leads to a distribution of speedy balls:

$$f(v) = (1/v)\ \exp(-v/v_o) \tag{1.a}$$

$$v_o = \beta b\omega/(\alpha n_l) \quad \text{(with } \alpha \leq 8, \beta=0.5-2) \tag{1.b}$$

$$n_l = N\pi d^2/S = Nd^2/D^2 \tag{1.c}$$

One sees that f(v) diverges as v tends to 0. But the hypotheses underlying the model are no more valid in this range [6], because collisions occur essentially with faster ball for instance, …; so the physics of this divergence has to be handled with care, and will be discussed a little later. As the number of impacts at speed v on a fix gauge varies as v f(v) and since the momentum transfer I is

$$I=mv(1+\varepsilon) \tag{2}$$

So, the distribution p(I) of momentum which corresponds to impacts with the gauge whose intensities are in the range I and I+δI shall be such that p(I) δI = vf(v) δv; so p(I) shall vary as :

$$p(I) \propto \exp(-I/I_o) \tag{3.a}$$

$$I_o = mv_o(1+\varepsilon) = I_o = m(1+\varepsilon)\ \beta b\omega/(\alpha n_l) \quad \text{(with } \alpha \leq 8, \beta=0.5-2) \tag{3.b}$$

*Remark 1:* If the *whole cell is moving sinusoidally*, instead of being fix and closed by a single moving piston, then the amplification process occurs twice a roundtrip leading to α≤4, instead of 8. In this case however, the impacts should be measured on an immobile plane perpendicular to z to correspond to the present modelling; hence these impacts do not correspond to the ones on the moving walls; nevertheless the correct distribution can be obtained from the impacts on a moving wall if the times at which they occur are known, using a correction algorithm for both the intensities and the rate of impacts as a function of the phase of the wall motion [6].

***Remark 2:*** The factor $\alpha \leq 8$ in Eqs. (1.b) & (3.b) are issued from the factor 4 from Eq. (1.c). This assumes that any ball-ball collision is equivalently efficient and leads to complete energy loss; this might be super evaluated since very eccentric impacts are likely less dissipative. Hence one expects that $\alpha < 8$.

***Remark 3:*** The model is non local, since it computes directly the probability of rare events which are issued from a long exploration of the cell, involving large trips with many roundtrips and many successive collisions with the pistons.

***Remark 4:*** The statistics the model predicts is strongly non Boltzmannian, as demonstrated by Eqs. (1) & (3), with a $1/v$ pre-factor and an $\exp(-v/v_o)$ tail.

***Remark 5:*** All the above physics is induced by the piston motion. This leads to a very peculiar trend which is non Boltzmannian; hence this leads to define a new kind of boundary condition, which acts for dissipative gas: the "velostat" boundary condition, or "impact generator" boundary condition [5, 6, 8].

***Remark 6:*** The role of few different effects have been discussed in [5,6] among which the possibility of a non-linear behaviour linking I to v (as in Hertzian contact), the possibility of a speed gain which depends on v during ball-piston collision, the effect of gravity, the effect of supersonic excitation,… .

***Remark 7:*** One notes by passing that Eq. (1) predicts the limits of ***supersonic excitation***, which imposes $v_o > b\omega$; it reads $n_l > 1/8$ for a vibrating piston and $n_l \gg \frac{1}{4}$ for a moving cell. This agrees with experimental finding approximately [9], since well established supersonic kind of excitation has been observed for $n_l=1$ in the case of the moving cell of MiniTexus 5 experiment, which proves it is already well above the condition of supersonic excitation. (Supersonic excitation is a concept which arises from physics of classic gas, for which it means that the typical ball speed, *i.e.* $v_o=\langle v \rangle$ is slower than the speed of the boundaries; this concept can be transposed to the dissipative case because one knows that the speed of sound c is related to the ball speed, since it is given by $c=\langle v \rangle (C_p/C_v)^{1/2}=v_o(C_p/C_v)^{1/2}$ with $C_p/C_v=7/5$ from classic gas theory of complex molecules).

***Remark 8:*** The model works in the limit of $b/L \ll 1$, otherwise a periodic regime can be reached rapidly, which writes $kvT/2 \approx L$, with $T=1/f=2\pi/\omega$ and k being a small positive integer.

***Remark 9:*** Applying a similar model with similar hypotheses on a granular layer in 1g with no lid leads to a time of flight $\tau_f=2v/g$. This modifies importantly the distribution, which reads $\mathbf{f_{1g}(v)= (Av/g) \exp(-v/v_o)}$, (when v is given at altitude z=0). This distribution is peaked on $v_o$, little balls have now small speed; the physics is then completely altered and looks much more as a normal gas. Further interaction between balls may even randomize more the system and make it looking more to a Boltzmann gas. The bottom boundary condition is also different [6] since it writes $g=\int v\, f(v)dv$. This may explain why granular gas in weightlessness is so specific. The simple fact that gravity forces slow particles to localise on bottom forces them to gain energy and to be reduced in number; it modifies then completely the physics. For instance, turning the cell with a single piston upside down in 1g leads to a condensate at v=0, instead of "gas", which demonstrates that the physics 1g and -1g is not at all the same.

This allows also asking in which amount g-jitter may alter experimental data in weightlessness.

## 2.  Comparison with experimental data :

With the support of CNES and ESA, we have performed a series of experiments in micro-gravity condition using the Airbus A300-0g of CNES, in the configuration using a fix cylindrical cell (Diameter D=13mm, length $L_o$=10mm) and a vibrating piston (b<5mm, 30Hz≤ω/(2π)≤120Hz, with balls ranging from 0.5mm to 5mm diameter. Data corresponding to impact intensities are reproduced in Fig. 1 in the case of 2mm diameter balls. The exponential behaviours of p(I) (Fig. 1) are well predicted by the above model; and the predicted scaling with bω and N are found . Anyway, the model catches likely the essential of the physics.

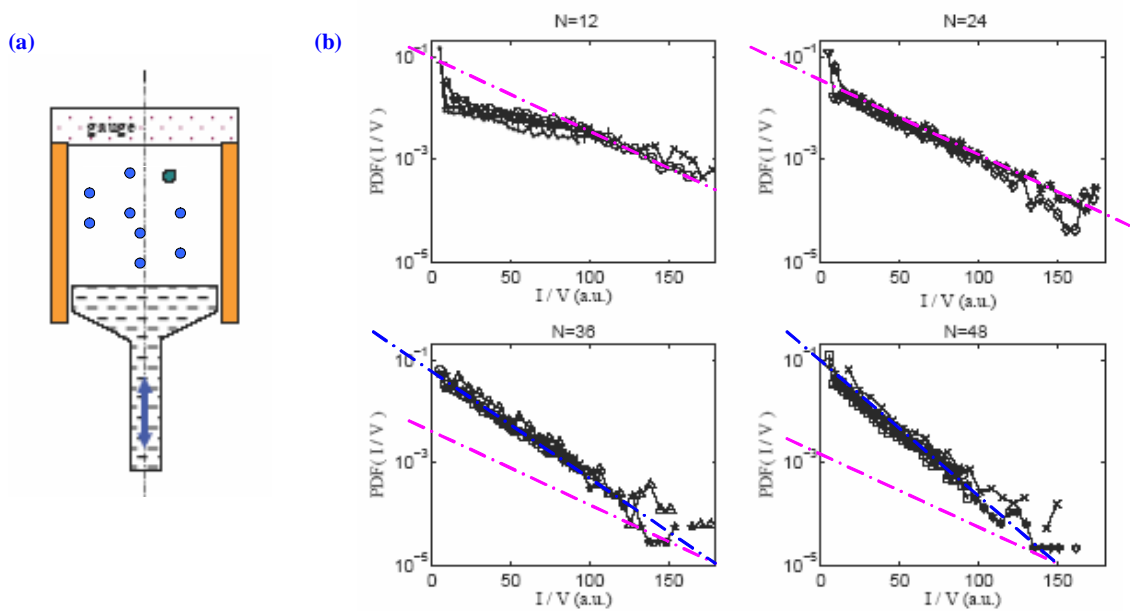

*Figure 1: (a, left): the experimental set-up which has been flown in Airbus A300-0g of CNES. (b, right) Probability density functions of the impact amplitude I measured by the sensor, for different vibration parameters during 16 s of low gravity, for different numbers of balls: N=12, N=24, N=36, N=48. V=bω. Symbols are N=12 [#1 (x); 2 (○); 3 (●); 4 (+)]; N=24 [#5 (\*); 6 (◊); 7 (  )]; N=36 [#8 (  ); 9 (pentagrams); 10 (○); 11 (x)] ; N = 48 [#12 (x); 13 (hexagrams); 14 (○); 15 (□)].One can rescale all the curves into a single one, using the parameter I $N^{0.8\pm0.2}$ /(bω), cf. [4,5,6].*

As a matter of fact, as discussed in [6], the response of the impact gauge has been calibrated during the same series of experiments, with a cell containing a single ball only [10, 11]; the response scales as I ∝ $v^{4/5}$, and its response time varies as $\tau_I$ ∝ $v^{1/5}$ , *i.e.* as for Hertzian contact law. This might explain some discrepancy between the theory and the experiments; in particular it can explain the $I_o$ ∝ $N^{-0.8}$ scaling reported in Fig. 1 caption, instead of the $v_o \propto N^{-1}$ predicted from Eq. (3), *cf.* [6]. Other reasons for slight deviations have been discussed in [5, 6].

Similar experiments have been repeated during Maxus 5 rocket flight in moving cells; one of the cell (of 1 x 1 x 1 cm³) was containing 50 balls ($n_1 \approx 0.4$) and was closed by a gauge. It has led to similar trends, *i.e.* $p(I) \approx \exp(-I/I_o)$ are approximately found from first-glance analysis of the data; however this needs to be confirmed because experimental data require further processing to be corrected from the cell motion.

As discussed in [9], data on pressure-fluctuations from MiniTexus 5 experiments cannot be used to infirm the above modelling, due to the lack of knowledge on the way that pressure gauge was working. (Indeed, this has led to define the new system of impact measurement, which has been tested and used in Airbus and used in Maxus 5).

### ■ *Distribution of ball speed:*

So, this exponential impact distribution is in accord with the speed distribution of Eq. (1). Hence it leads to a divergence of its integral, $\int f(v) \, dv$, due to the slow-speed contribution. This divergence may be avoided by introducing a cut-of speed $v_1 \ll v_o$ as a lower limit of integration. Nevertheless, it tells that most of the balls are very slow and that only some are moving fast. Also, the impacts generated by very slow speed are small amplitude; so they are hardly detectable. Hence one cannot tell exactly what the effective experimental distribution at very slow speed is. There might be more such balls for instance. In particular, one sees a peak which appears in the distribution in the v=0 range in Figs. 1 corresponding to N=12 and N=24. Is this demonstrating the existence of an anomalous distribution at slow speed?

### ■ *Mean ball speed:*

On the other hand, the mean ball speed $<v_1>$ can be defined as

$$v_m = \{\int v \, f(v) \, dv\} / \{\int f(v) \, dv\} = v_o / \{\int f(v) \, dv\} \tag{4}$$

Hence it depends on $\int f(v) \, dv$, which is quite sensitive to the limit $v_1$.

On the other hand, $v \, f(v) = p(I)$; so the impact distribution $p(I)$ of Fig. (1) characterises an assembly of balls with typical speed $v_o$. The very small impacts, if they exist, do not seem to be important. So, a first way to treat the problem is to normalise the distribution $\{\int_{v_1}^{\infty} f(v) \, dv\} = 1$. Hence we get $v_m = v_o$.

There is also an other way to measure $v_m$: One just needs to count the total number $N_c$ of impacts on the gauge during a time T. This shall scale as

$$N_c = N \, v_m T / (2L) \tag{5}$$

Then if $v_m = v_o$, Eqs. (1) & (5) imply $N_c = N \, T \, v_o = N \, T \, \beta b \omega / (\alpha n_l 2L)$, with $n_l = Nd^2/D^2$; so one expects:

$$N_c = T \, D^2 \, \beta b \omega / (\alpha d^2 2L) \tag{6.a}$$

which can be written as

$\quad N_c = (D/d)^2 \ (\beta/\alpha) \ b\omega T/(2L)$ (6.a)

As one has $0.5<\beta<2$ and $3<\alpha<8$, the expected range of $(\beta/\alpha)$ is 0.06-2/3; a good estimate is expected around $\beta/\alpha \approx \frac{1}{4}$ which corresponds to $(\beta=2, \alpha=8)$ or $(\beta=1, \alpha=4)$. Then using numerical values (D=13mm, d=2mm, $\beta/\alpha = \frac{1}{4}$, T=16s, 2L=16mm) leads to $N_{c,T=16}$:

$\quad N_c = N_{c,\ T=16} \ b\omega$ (6.b)

with $N_{c,\ T=16} = (D/d)^2 \ (\beta/\alpha) T/(2L) = 42250 \ (\beta/\alpha) \approx 10560$ (6.c)

so $\quad N_{c,\ T=16} \approx 10560$ if $\beta/\alpha = \frac{1}{4}$ or $\quad N_{c,\ T=16} \approx 21000$ if $\beta/\alpha = \frac{1}{2}$ (6.d)

| # | N | b | f | V=bω | Γ | $N_c$ | <v>/(bω) | $N_{c,T}$= |
|---|---|---|---|---|---|---|---|---|
|   |   | mm | Hz | m/s | bω²/g |   | (L=8mm) | $N_c/(b\omega)$ |
| 1 | 12 | 0,92 | 40 | 0,23 | 5,9 | 2591 | 0,94 | **11265** |
| 2 | 12 | 0,65 | 59,7 | 0,24 | 9,3 | 2605 | 0,90 | **10854** |
| 3 | 12 | 0,88 | 80 | 0,44 | 22,7 | 5735 | 1,09 | **13034** |
| 4 | 12 | 0,64 | 90,9 | 0,37 | 21,4 | 4617 | 1,04 | **12478** |
| 5 | 24 | 0,96 | 40 | 0,24 | 6,2 | 5097 | 0,89 | 21238 |
| 6 | 24 | 0,67 | 59,7 | 0,25 | 9,6 | 4078 | 0,68 | 16312 |
| 7 | 24 | 0,88 | 80 | 0,44 | 22,8 | 8362 | 0,79 | 19005 |
| 8 | 36 | 0,44 | 40 | 0,11 | 2,8 | 2538 | 0,64 | 23073 |
| 9 | 36 | 0,67 | 59,7 | 0,25 | 9,7 | 6496 | 0,72 | 25984 |
| 10 | 36 | 0,88 | 80 | 0,44 | 22,8 | 9744 | 0,62 | 22145 |
| 11 | 36 | 0,69 | 90,9 | 0,39 | 22,9 | 9741 | 0,70 | 24977 |
| 12 | 48 | 0,42 | 40 | 0,11 | 2,7 | 2728 | 0,52 | 24800 |
| 13 | 48 | 0,69 | 59,7 | 0,26 | 9,9 | 8650 | 0,70 | 33269 |
| 14 | 48 | 0,89 | 80 | 0,45 | 22,9 | 10906 | 0,50 | 24236 |
| 15 | 48 | 0,73 | 90,9 | 0,41 | 24,2 | 12512 | 0,63 | 30517 |

**Table 1: Number $N_c$ of collisions during T=16s for a cell containing N balls (N=12, 24, 36, 48), for different amplitude (b) and frequency f=(ω/2π) . Last column corresponds to experimental value of $N_c/(b\omega)$.** They shall be compared to $N_{c,T=16}$ estimated from Eq. (6.c): $N_{c,T=16}$= (D/d)² (β/α)T/(2L) = 42250 (β/α ) (≈10560 for β/α=¼). Most probable range for β/α: 0.05<β/α<0.7.
Data on <v>/(bω)=$N_c$L/ (TπbfN) are 20% less than in Table 1 of [1] because the L=8mm value is corrected from the ball diameter here (and was not there).

Hence Eq. 6 tells that the number $N_c$ of collisions should scale proportional to the excitation maximum speed V=bω and to be independent of N, and predicts the number of counts. All the data of Table 1 falls in the prediction within the precision error bar. However, it shows that $N_c$ depend on N. Another point which is demonstrated by Table 1 is the fact that error bars on $N_c$ is larger than the normal error bar $(N_c)^{\frac{1}{2}}$ as one should expect from counting random events. This might be due to g-jitter, or indicate 10-20% accuracy on bω, which is rather large. It might also indicate some special cooperative effect which increases the noise and which remains to be determined and studied. Next part of the section does not take into account these possible explanations and neglect these effects; it might be an error (*cf.* remark 9 of section 1).

Another way of writing Eq. (6) is to evaluate the number $N_{c,1/f}$ of collisions per period of excitation; calling S the surface of the top, one gets:

$$N_{c,1/f,piston}= (S/d^2)(\beta/\alpha_{piston})(b/L) \qquad (7.a)$$

$0.5<\beta<2$ ; $1< \alpha_{piston} <8$ ; d=ball diameter, S=surface of cell section

*Remark:* When comparing the problem of Knudsen granular gas in a vibrating cell to the one of a granular gas excited by a piston in a closed cell, similar trend is expected. However, one shall take into account that similar amplification occurs twice in a period so that dissipation due to ball-ball collisions is controlled by $l_c/L$ instead of $l_c/(2L)$; this implies $\alpha_{cell}= \alpha_{piston}/2$. This introduces the factor 2 between Eqs. (7.a) & (7.b).

$$N_{c,1/f,vibrated-box}= (S/d^2)(\beta/\alpha_{cell})(b/L) \approx 2\, N_{c,1/f,piston} \qquad (7.b)$$

### ■ One can define two typical mean speeds at the same time!

An other way to look at the same data is presented in Fig. 2, which reports the variation of $N_c$ vs. $b\omega$ for a time T=16s of recording and for different number N of balls in the cell. Any detectable impact has been included (as in table 1). Fig. 2 shows also that $N_c$ varies linearly with $V=b\omega$, as expected, but it demonstrates that $N_c$ depends also on N and seems to obey $N_c \propto N^{0.6}$. (Nevertheless, the range over which the scaling has been studied, *i.e.* N=12-48, is too small, so that the scaling law is not proved; it is just a commodity of writing). In other words, since $N_c=Nv_m T/(2L)$, this result leads to a mean speed $v_m$ that scales as :

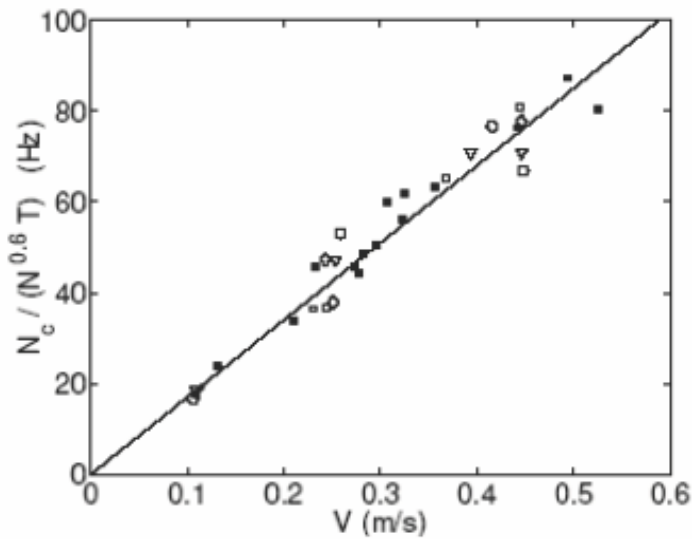

*Figure 2: Total number of collisions $N_c$ observed during experimental time T=16s, rescaled by $N^{0.6}$ and by time, $N_c/(T\, N^{0.6})$, as a function of $V=b\omega$ for N =12 balls (□) and (■); 24 balls (♦); 36 balls (∇); 48 balls (o) . T = 16 s is the duration of steady low gravity in Airbus A300-0g of CNES. ■ marks are from experiments with N=12 balls and 15 different velocities (cf. [1]). From [4].*

*N is the number of particles in the cell. Solid line corresponds to the fit $N_c/(T\, N^{0.6}) = \alpha V$.*

$$v_m \approx b\omega/N^{0.4} \qquad (8)$$

So, $v_m$ is obviously different from $v_o$, since this one varies as 1/N according to Eq. (1.b). On the other hand, the experimental distributions in Fig. 1 define also a set of typical speeds; but this one obeys Eq. (1.b) and it is found to decrease as 1/N at least in first approximation. So, it seems that there are two different ways of estimating a mean speed <v>, which lead to two different trends, either in $<v>=v_m \propto 1/N^{0.4}$ or in $<v'>=v_o \propto 1/N$. Both are related to the Oz component of speed only. Is there some experimental mistakes?

In fact one can perhaps understand this apparent incompatibility within the following argument: on one hand, owing to the $(A/v) \exp(-v/v_o)$ distribution, the probability of getting a speedy balls is very little, although they contribute much to the impact statistics. So most of the balls are merely at rest; and they collide the piston more often than the speedy balls. On the other hand the main way these slow balls can gain some energy is not from a collision with the piston but from collision with a fast ball, because their speed v is much slower than bω, *i.e.* $v \ll b\omega$, so that their collision with the piston occurs mainly when this one is slow, *i.e.* near its maximum elongation. It results from this that the probability of escaping from such a state is smaller from the others; this may enhance likely the proportion of such slow balls. Within this view point, one shall conclude the distribution may be two fold; the first category consists of balls merely at rest whose distribution $f_r(v)$ is peaked and centred around $v_r \ll v_o$; and the second one is made of the speedy balls whose statistics obey $f_g(v)=(A/v) \exp(-v/v_o)$ at large speed. Both communities interact through collisions and form a dynamical equilibrium.

This seems to be verified experimentally indeed, since one sees a peak of distribution near v=0 in Fig. 1 for N=12 and for N=24. This peak is washed out after, for N=36 and 48. At the same time, one sees the increase of the ordinate at the origin of the exponential tails, when passing from Fig. 1 N=12→24→36→ 48. It means that the number of balls that play the game of gaining some speed per roundtrip increases from N=12→24→36→ 48; conversely, it means that the number of particles almost at rest decreases from N=12→24→36→ 48, and the proportion of speedy ball increases. This makes $v_m$ increasing faster than $v_o$ in this range. At last $v_o$ is expected to follow Eq. (1) since the probability of important loss is related to the collision probability with any of the two kinds of balls (speedy or slow) which depends still on $L/l_c$, with $l_c$ independent of v and depends on the total bead number.

So the proposed modelling tells that there shall be equilibrium between two kinds of balls:
 (i) the ones which are almost at rest, whose number is $N_r$, with a distribution peaked around $v_r$ ($v_r \ll b\omega$), generating $N_{cr}$ collisions with the gauge at a typical speed $v_r \ll b\omega$, which leads to $N_{cr}=N_r v_r T/(2L)$.
 (ii) the balls whose speed is amplified "normally" by the piston, whose characteristic speed is $v_o$ given by Eq. (1), whose number is $N_g$; these balls generate a distribution $p_g(mv)$ of impacts that follows $p_g(mv)= A \exp[-mv/(mv_o)]$, and $N_{cg}$ impacts, such as $N_{cg} \approx A v_o T/(2L)$ and $N_g = \int_{v_1}^{\infty} (A/v) \exp(-v/v_o)dv$.

This leads to the following set of equations :

$$N_r + N_g = N \tag{9.a}$$

$$v_o = \beta b \omega / (\alpha n_l) \quad \text{with } \alpha \leq 8, \beta = 0.5\text{-to-}2 \text{ and } n_l = N\pi d^2/S = Nd^2/D^2 \tag{9.b}$$

$$N_c v_m = (N_{cr} + N_{cg}) \, 2L/T \tag{9.c}$$

$$N_r = N_{cr} \, 2L/(v_r T) \tag{9.d}$$

$$N_g = N_{cg} \{2L/(v_o T)\} \int_{v_1}^{\infty} (1/v) \exp(-v/v_o) dv \tag{9.e}$$

Eq. (9.e) allows determining the speed $v_1$ in order that Eqs. (9.a) and (9.d) are satisfied at the same time. This fixes the amount of speedy ball $N_g$. Neither $N_r$ nor $N_g$ can overpass $N_c$; this fixes some limit.

In theory the problem has a solution. However, it can be unphysical; for instance, it becomes unphysical if $N_{cr}$ is large and $v_r$ is small enough so that $N_{cr} \, 2L/(v_r T) \gg N$.

***Effect of heterogeneity:*** As noted before, slow balls are hardly excited by direct collision with the piston because this occurs when the piston is at maximum elongation; so there might be some excess of slow balls. This generates cases for which $N_g < N$. The probability of de-exciting a fast ball remains linked to the probability of its collision with any other ball; hence it remains the same as far as all the other balls (either fast or merely at rest) are randomly distributed; this let predicts a $v_o$ speed which scales as $1/N$.

However, slow balls can arrange spatially due to (i) mutual interaction, or (ii) to interaction with the walls, or (iii) interaction with balls in the gas; this is expected to occur when clustering forms or if the balls with little speed organise themselves in lines parallel to vibration, or are stuck on walls by electro-static attraction,…; another case might be also the appearance of some natural hierarchy of heterogeneity, such a hierarchy of heterogeneity exists in the universe where one sees, stars, clusters of stars, galaxies, cluster of galaxies…. So it might occur that collision between balls from the "gas" and balls in the phase "nearly at rest" be rarer than normal. In this case one expects $v_o$ to be larger than predicted.

***Pressure:*** It has been shown that the impacts of the gas phase on a fix gauge are dominated by slow balls, owing to their numbers $f_g(v) = (A/v) \exp(-v/v_o)$ that leads to $p_g(I=mv) = A \exp(-I/I_o)$. However, the pressure $P_g$ is produced mainly by faster balls according to $P_g = \int v^2 f(v) dv \approx A v_o^2$.

## 3. Comparaison with experiment on Maxwell's demon:

In the experiment on Maxwell's demon one investigates the equilibrium between a system of two halve boxes partly filled with balls and connected through a horizontal slit. The system is vibrated vertically. If the number of ball layers is not too large, one observes the equi-partition of the balls above a threshold of vibration; but the distribution becomes biased below this threshold, and the bias depends on the distance to the threshold. At last the threshold is found to depend on the total number of balls.

A way to understand the phenomenon has been to consider a single half box, and to study the way it empties with time and with the intensity of vibration, measured in acceleration unit. Fig. 3 reports a typical result [12]; the number of layer is counted with reference to the hexagonal distribution (so such a layer contains 0.9 the number of balls defined in a layer obeying to Eq. 1.c). One sees in Fig. 3 that the number of balls which passes through the slit per unit of time increases continuously with $n_1$ till a maximum, then it decreases. This maximum occurs at $n_{1max}$ which depends on the acceleration, but which is smaller than ½ .

It has been shown in [12] that bias distribution occurs in a system of two halve-boxes connected by a slit when the mean number of layer overpasses $n_{1,max}$.

On the other hand if the slit was closed and replaced by a gauge, the gauge will count a collision each time a ball tends to escape. Hence $j_\perp = dN/dt$ can be viewed also as the number of collisions with the slit, during a fix time. It is then the analogue of $N_c$, but measured perpendicularly from the vibration. We will note both flows $j_\perp$ and $j_{//}$ respectively.

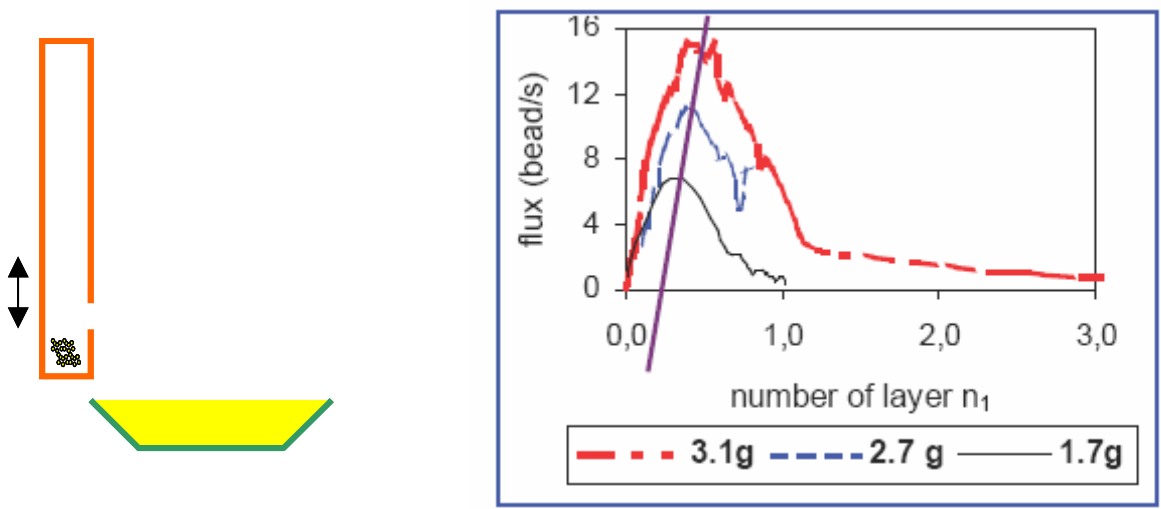

*Figure 3: 1-g experiment on Maxwell demon, studied using the emptying of a half box with no lid on top:*
*Left: the experimental set-up. Right: Flow $j_\perp=dN(t)/dt$ of grains flowing from the first compartment as a function of N of grains remaining in the compartment #1, expressed in term of the number $n_1$ of layers in compartment #1 (right Fig.), at different accelerations Γ, Γ=2.7g, 3.1g and 3.4g. It is worth noting that (i) j varies linealy with N (or $n_1$) at small N, (ii) that $j_\perp$ exhibits a maximum, which occurs much before n= ½ , and (iii) that non linear behaviour occurs at n ≥ ¼ . ( from Fig.6 of [12]).*
*$n_1$ is measured here as in reference [12], i.e. as $n_1=1$ for the dense hexagonal packing; hence it is 0.9 the value of a layer defined by Eq. (1.c) .*

As $j_{//}$ is found to increase in between $n_1$=0.3-1.3 according to $N_c \propto j_{//} \propto N^{0.6} \propto n_1^{0.6}$, *cf.* Fig.2, the result of Fig. 3 demonstrates probably that the maximum of $j_\perp$ occurs at much smaller $n_1$ than $j_{//}$. Such a difference indicates a strong difference in between the two processes and demonstrates that granular gas exhibits some important anisotropy.

The reason is likely simple: to go out from a wall perpendicular to vibration the balls need better not to collide with other balls, so that direct flights allow much more efficient escaping in this configuration. On the other hand, escaping from a lateral wall requires a collision with another ball, just to get the right direction of speed, hence $j_\perp$ is linked to the speed distribution after a collision with a ball, while $j_{//}$ is related to the one before a collision. Also a screening effect can reduce the contribution to $j_\perp$ ; it is generated when the mean free path is small: for instance, the efficiency of collisions which occur in the cell centre is reduced importantly when the mean free path $l_c$ becomes smaller than the cell width $L_w$. Assuming a cell with cubic shape $L^3$, taking $l_c = L/(4n_l)$ and $l_c = L/2$ or L as a limit to find an efficient screening, one predicts $n_{1,max}=$ ½ or ¼ as the typical threshold for which the screening occurs.

$$n_{1,max}= ½ \text{ or } ¼ \tag{10}$$

This is just the position of the maximum of $j_\perp$ observed experimentally in Fig.3.

Although this explanation sounds satisfying, it requires also further investigation, to measure in particular the true effect of gravity. Indeed, as shown is section 1, remark 9, the behaviour of granular gas in 0g is expected to be quite different from in 1g. Nevertheless, the present section demonstrates also some connection between the two physics.

## 4. Conclusion:

This article considers a granular dissipative gas in a Knudsen regime and excited by a vibrating piston, for which it proposes a modelling of the speed distribution. It is based on a mechanical analysis of the problem in which the piston plays the role of a random impact generator, or of a "velostat" (speed generator) and where the granular "gas" plays the role of a random dissipater. So the problem is settled in terms of probability, energy gain and energy losses during collisions with piston and with the balls respectively. It takes into account the true role of boundaries and the role of dissipation of the gas. It predicts the exponential decay of p(I) *vs*. I, where p(I) is the probability of finding an impact of size I $\propto$ mv, when the gain of speed per piston speed is independent of v. This leads also to the speed distribution f(v)= (A/v) exp-v/v$_o$), where v$_o$ is the typical speed of the distribution.

The model predicts the right variations of
(i) v$_o$ *vs.* n$_1$, the number of layer covering the bottom of the cell at rest.
(ii) v$_o$ *vs.* bω , the piston speed

It predicts also approximately the correct amount N$_c$ of collisions with a fix gauge. However this one is found experimentally to depend slightly on the number of balls contained in the cell while the models predicts no dependence. The misfit is attributed to balls "merely at rest" which are in excess; hence it is equivalent to introducing a second phase which describes this set of balls at rest. A reason for this excess is described; it comes from the fact that very slow balls do not gain much energy when they collide with the piston, because their collision occurs when piston is at maximum elongation. $N_c$ is found to vary linearly with the piston speed $b\omega$, and with S/L, the ratio of the gauge surface to the cell length $L=L_o-d$.

The effect of the second phase is discussed as a function of its organisation: if it is random, $v_o$ remains the same; but a clustering of this phase may reduce the probability of collision between a speedy ball and this phase, which favours the existence of such speedy balls and increase $v_o$. However this might occur only in very large cells.

As demonstrated in remark 9 of part 1, the model leads to completely different prediction in 1g, with no accumulation of balls having a slow speed. It may occur also that the models be affected by g-jitter; in particular, it is possible that the number of particles which are merely at rest depends importantly on this parameter.

The present theory investigates the speed distribution in the direction of vibration; but some other experiments are more interested in transverse motion such as the Maxwell's demon experiment. Some understanding can be gained from their confrontation, and the compatibility of their results can be tested. In Maxwell's demon experiment it has been found that the transverse flow $j_\perp$ passes by a maximum for $0.3<n_{1,max}<0.5$, while the longitudinal flow $j_{//}$ of the present experiment is still increasing for $n_1>1$. This difference is described in term of the processes: $j_{//}$ is controlled by the speedy particles moving parallel to the excitation; its main contribution comes from speedy particles which do not meet slow particles. On the contrary, contribution to $j_\perp$ comes from fast particles which hit other particles to get moving laterally; after this primary collision, such trajectories feel some screening effect due to secondary collisions; this reduces the contribution to $j_\perp$ from collisions which occurs in the cell centre when the mean free path $l_c$ becomes smaller than the cell width $L_w$. We have found that such modelling predicts a maximum of $j_\perp$ at $n_{1,max}<0.5$ about; so it describes the correct value of $n_{1,max}$. However, further experimental and theoretical investigation is required to confirm this analysis.

**As a conclusion:** I do not believe that the trends and ideas developed in the present article were the current belief at the moment. Furthermore, the $\exp(-v/v_o)$ was not observed by simulations, nor predicted by theories. And the correct speed distribution, $f(v)= (A/v) \exp(-v/v_o)$, is merely "incredible". One shall ask why simulations did not get that trend, since the experiments we used are not difficult to simulate, nor they need large number of balls? Are our experimental results wrong? I do not believe since Maxus 5 experiments seem to confirm the trend. So it reveals probably some important error in the protocol when using simulations alone, or simulations coupled to theory or to experiments.

For instance, experimental protocole tells how to find errors, incompatibility between results …; for instance, the experiment with 1 ball has been performed to calibrate a gauge, but its main result has been to demonstrate the important dissipation which exists during collision due to the coupling between translation and rotation degrees of freedom [10-11]. This is then a typical by-product which is important and probably more difficult to generate using computers.

So, thanks to CNES and ESA, the series of micro-gravity experiment tends to prove the following facts:
- strong effect of dissipation due to translation-rotation coupling during collision (this is mainly neglected in simulations), see [5,10,11]
- An anomalous distribution of speed in the Knudsen regime
- A non local theory is needed to explain the data of speed distributions, with an amplification factor due to interaction with boundaries, and a loss term to collision with gas.
- This non local theory predicts the supersonic character of the excitation for low density systems $n_1 > 1/4$, which is observed experimentally.
- Experiment shows also the existence of a rather 2-phase system: the first phase is not really moving, it is made of a cloud of particle "at rest", located in the middle of the cell and disconnected from boundaries. The second phase is made of particles moving back and forth in between the piston and the lid, or in between the two moving walls (depending on which experimental set-ups is considered). So the second phase interact with boundaries and is slowed down by interaction with the cloud (1$^{st}$ phase); hence it maintains also the cloud (1$^{st}$ phase) in the middle of the cell. This confirms the idea of [9].
- These trends do not appear (or do not appear important) from the reading of works from other teams. [1-3]. Our results are mainly not quoted too, even if they have been published (see list below), and defended in different conferences; they seem to be known but they are probably considered as different negligible results and thinking [1] at the stage of scientific debate.

It seems that the present behaviour shall be modified by gravity and/or g-jitter; this is a new orientation of research. Also, the case of much denser samples is out from the scope of the present theory. One may ask for instance, if the gas phase, which is defined as the phase whose speed distribution is $f(v) = (1/v) \exp(-v/v_o)$, is equivalent to a skin effect which disconnect the inner part to the boundary condition. This cannot be true according to the present modelling which shows that the balls in the gas phase perform permanent roundtrip from a boundary to the opposite one. So this gas phase fills up and sounds the complete cell. The faster the ball is, the larger the path in "vacuum" and the more numerous the interaction with boundaries only, so the more "Knudsen" the particle. Any particle looks moving slowly for along time, then it hits a speedy ball or a moving wall and starts moving faster and faster as it collides successively the moving walls, till it hits a slow ball, and becomes quiet. So the dynamics of each ball appears as short bursts; but, the global dynamics is more random because balls in the cell are numerous and their motion not correlated. The large noise observed on the number of collisions $N_c$ is perhaps due to the existence of the bursts which contain few collisions, although it can be induced also by g-jitter.

When considering the case of an important amount of second phase whose typical ball speed is much smaller, it can be viewed in some case as rather decoupled to the "gas" phase. This might be true and observed; however this second phase can be coupled also to the boundary, via a fluid such as the air for instance; (if air fills up the cell,…). So the phase "at rest" may move also periodically, excited by the motion of the boundaries. This is a new way for the granular medium to be connected to the wall dynamics, which is not investigated in the present paper. But it will occur likely at larger number of layers. The speed distribution which is generated by such a coupling will be quite different from the present one; one has not to confuse them, because the dynamics with a fluid will penetrate much deeper in the cell and generate internal flows. But this is a new full topic [13,14].

*Acknowledgements:* CNES and ESA are thanked for they strong support and for funding the series of parabolic flights in board of the Airbus A300-0g. Assistance from Novespace team has been quite appreciated. The experimental results have been obtained by the team composed of D. Beysens, E. Falcon, S. Fauve, Y. Garrabos, C. Lecoutre, F. Palencia and myself. They are all thanked.

and have been explained in a series of conferences

The electronic arXiv.org version of this paper has been settled during a stay at the Kavli Institute of Theoretical Physics of the University of California at Santa Barbara (KITP-UCSB), in june 2005, supported in part by the National Science Fundation under Grant n° PHY99-07949.

*Poudres & Grains* can be found at :
http://www.mssmat.ecp.fr/rubrique.php3?id_rubrique=402